%% file: main.tex
\documentclass[submission,copyright,creativecommons]{eptcs}
\providecommand{\event}{GCM 2025} 

\usepackage{graphicx}
\usepackage{iftex}
\usepackage{amssymb,amsmath}
\usepackage{stmaryrd}
\usepackage{stmaryrd}
\usepackage{pgf, tikz}
\usepackage{algorithm}
\usepackage{algorithmic}
\usepackage{fancyvrb}
\usepackage{mdframed}
\usepackage{pgfplotstable}
\usepackage{pgfplots}
\usepackage{enumitem}
\usepackage{chngcntr}
\usepackage{capt-of}
\usepackage{cite}
\usepackage{color, colortbl}

\ifpdf
  \usepackage{underscore}         
  \usepackage[T1]{fontenc}        
\else
  \usepackage{breakurl}           
\fi

\title{Implementing Binary Search Trees in GP\,2 \\ (Extended Abstract)}
\author{Ziad Ismaili Alaoui
\institute{University of Liverpool, United Kingdom}
\and
Detlef Plump
\institute{University of York, United Kingdom}
}
\def\titlerunning{Implementing Binary Search Trees in GP\,2}
\def\authorrunning{Z. Ismaili Alaoui and D. Plump}
\begin{document}
\maketitle


\vspace*{-1ex}
\section{Introduction}
We present an approach to implement binary search trees in the rule-based graph programming language GP\,2. (See \cite{campbell2022fast} for a brief introduction to GP\,2.) Our implementation uses GP\,2's rooted graph transformation rules to be fast \cite{Alaoui-Plump25b} and supports insertion, deletion and query operations. We argue that the worst-case runtime for each of the operations is $\mathrm{O}(n)$ for a tree with $n$ nodes. In addition, we expect that, on average, the operations run in time $\mathrm{O}(\log n)$\footnote{Here and throughout, $\log = \log_2$.}. Hence the implementation would match the time complexity of binary search trees implementations in imperative languages (see, for example, \cite{Skiena20a}). 

\section{Binary Search Trees}
\label{s:bst}

\begin{figure}[!ht]
    \centering
    \begin{tikzpicture}
        \node[shape=circle,draw=black] (1) at (0,0) {$5$};
        \node[shape=circle,draw=black] (2) at (-1,-1) {$2$};
        \node[shape=circle,draw=black] (3) at (1,-1) {$7$};
        \node[shape=circle,draw=black] (4) at (-2,-2) {$1$};
        \node[shape=circle,draw=black] (5) at (0,-2) {$4$};
        \node[shape=circle,draw=black] (6) at (2,-2) {$8$};
        \path [very thick, ->] (1) edge (2);
        \path [very thick, ->] (1) edge (3);
        \path [very thick, ->] (2) edge (4);
        \path [very thick, ->] (2) edge (5);
        \path [very thick, ->] (3) edge (6);
    \end{tikzpicture}
    \caption{Example of a binary search tree.}
    \label{fig:bst-example}
\end{figure}
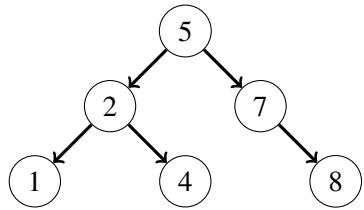

A \emph{binary search tree} (BST) is a binary tree in which each node is labelled with a distinct key (i.e. an integer value). For every node $v$, all keys in its left subtree are strictly smaller than the key of $v$, while all keys in its right subtree are strictly greater than the key of $v$.\footnote{Note this implies that BSTs do not have repeated entries.} Figure~\ref{fig:bst-example} shows an example of a binary search tree with $6$ nodes.
\noindent Binary search trees typically allow for three operations: insertion, querying, and deletion. We provide a high-level explanation as to how these operations work.

\paragraph{Insertion.} Inserting a node into a BST must preserve its properties. To insert a new key, we begin at the root and recursively traverse the tree: at each node, we move to the left child if the new key is less than or equal to the current node’s key, or to the right child if it is greater. This process continues until we reach a null child position, where the new node is inserted. This ensures that the BST property remains valid after insertion.

\paragraph{Querying.} Searching for a key in a BST proceeds similarly to insertion. Starting at the root, we recursively compare the target key with the current node’s key: if they match, the search is successful; if the target key is smaller, we search the left subtree; if larger, the right subtree. The search continues until the key is found or the search cannot proceed farther, indicating that the key is not present in the tree.

\paragraph{Deletion. } Similarly to insertion, deleting a node from a BST must also preserve its properties. There are three possible cases to handle:\\ 
(1) If the node to be deleted is a leaf, it can simply be removed, leaving its parent with one less child.\\
(2) If it has one child, the node is removed and its child is connected directly to its parent.\\
(3) If it has two children, its key is replaced with the maximum key in its left subtree, and that replacement node (which always is a leaf or has one child) is then deleted. To find the maximum key in the left subtree, we start at the left child and repeatedly move to the right child until no further right child exists; this also ensures that the replacement node does not have two children, in case of duplicate elements. This operation naturally preserves the ordering of the elements of the tree.\\

In a conventional programming language, these operations can be implemented to run in time $\mathrm{O}(n)$ in the worst case for a BST with $n$ nodes. (This is when the tree degenerates into a linked list.) If the BST is \emph{balanced}, that is, when the heights of the child subtrees of any node only differ by at most 1, these operations run in time $\mathrm{O}(\log n)$ (see \cite{Skiena20a}). 

\section{The Program}

The programming language GP\,2 uses the notion of so-called \emph{roots}, which are distinguished nodes that are stored independently from non-rooted ones. To avoid confusion, in this section, we refer to the root of the binary tree (i.e. the top element) as the \emph{top}. 

The program \texttt{bst} (Figures~\ref{fig:bst-prog} and~\ref{fig:bst-prog2}) implements a binary search tree in GP\,2 whose top is pointed at by a green node, which acts as a parent to the top, and follows the specification below.

\begin{description}
\item[\textbf{Input:}] A linked list of unmarked nodes and edges such that
\begin{enumerate}
        \item each node label represents exactly one user instruction (insertion, query and deletion);
        \item nodes labelled \texttt{"i":}$n$ for some \texttt{int} $n$ represent an instruction to insert $n$ into the BST; 
        \item nodes labelled \texttt{"s":}$n$ for some \texttt{int} $n$ represent an instruction to search for a node labelled $n$, should one exist, in the BST and create a dashed edge from the instruction node to the found node in the BST;
        \item nodes labelled \texttt{"d":}$n$ for some \texttt{int} $n$ represent an instruction to turn a node labelled $n$ in the BST, should one exist, into a garbage node\footnote{A \emph{garbage node} is a non-instruction node that is not reachable from the green node.};
        \item the first instruction is the head of the linked list and rooted; and
        \item there exists an edge from $u$ to $v$ if and only if $u$ denotes the $i$th instruction and $v$, the $(i+1)$th.
    \end{enumerate}
\item[\textbf{Output:}] A graph consisting of the linked list provided by the user, where the tail is rooted instead, a green node, and a binary search tree whose nodes are grey, edges are unmarked, and whose top is pointed at by the green node and constructed in accordance with the order of instructions specified by the user and the procedures described in Section~\ref{s:bst}. 
\end{description}

\input{program}

We refer to the nodes specified by the user as \emph{instruction nodes}. The program first applies the rule \texttt{make\_root} to create a green dummy node whose function is to serve as a parent to the top of the BST; this technique allows for simpler rule design. The program iterates through the user-given instructions by performing a traversal of the linked list provided as input. Throughout the execution of the program, an instruction node whose instruction is currently in execution is rooted; otherwise, it is not. We describe the behaviour of the program for each operation, modularised as procedures (\texttt{Insert}, \texttt{Search} and \texttt{Delete}), below.

\paragraph{Insertion (\texttt{Insert}).} The program first attempts to root the top of the BST by calling the rule \texttt{root}. If none exists (i.e. the tree is empty), the rule fails to apply and \texttt{add_root} is applied instead, which creates a top to the tree. If \texttt{root} successfully applies, meaning that at least one node is in the tree, the rule set \texttt{\{go\_right1, go\_left1\}} is called as long as possible.\footnote{The edges of the binary search tree need not be explicitly labelled, as each edge's direction (left or right) can be inferred directly by comparing the values of its endpoint nodes. This does not affect the complexity stated in Claim~\ref{c:complexity}.} Essentially, each rule moves the root down the tree until none can apply, which happens if and only if the root reaches a node of strictly less than two children, and the new key in the rooted instruction node can be added as a child to the rooted node in the BST. The insertion is done by the application of the rule \texttt{add\_leaf}. Given that the BST cannot hold duplicate elements, the rule \texttt{match} is tried each time the root is moved at some position and terminates the loop as soon as a node of existing key exists in the tree.

\paragraph{Querying (\texttt{Search}).} This procedure works in a way analogous to \texttt{Insert}. The program first attempts to root the top of the BST by calling \texttt{root}. If that rule fails, the procedure terminates as there are no nodes in the tree. The traversal remains similar; as soon as none of the rules in the rule set \texttt{\{go\_right1, go\_left1\}} is applicable, either the element was found or a dead end was reached. The rule \texttt{match} is then called and creates a dashed edge from the rooted instruction node to the rooted node in the BST holding the sought-for key should they match.

\paragraph{Deletion (\texttt{Delete}).} Similarly, an application of the rule \texttt{root} is attempted; a failure simply terminates the procedure as there would be no node to delete. Then, the node holding the key to be delete, should it exist, is rooted by performing a search analogous to that of previous procedures down the tree. Should that node be found, three different cases of deletion apply, as described in Section~\ref{s:bst}. If the node to be deleted is a leaf, the rule \texttt{delete\_leaf} applies. If it is a one-child node, the rule \texttt{delete\_midl} applies. Otherwise (i.e. the node has two children), the rule \texttt{save\_node} is applied to keep track of that node; then, the root is moved to the left child, and go downward to the right child as long as possible. Eventually, the rooted node in the tree contains the largest key of the child left subtree (LKCLS) of the node to be deleted. The keys are swapped by reconnecting the edges; this is done by the rules prefixed \texttt{swap} (there are $6$ distinct cases depending on many levels down the LKCLS is), and one of the other two cases now applies to the node to be deleted. Note that nodes are not actually deleted from the tree; indeed, they only become disconnected from the BST, that is, garbage nodes. The insertion of a previously deleted key does not affect garbage nodes (i.e. there could be multiple garbage nodes of the same key).

\newtheorem{claim}{Claim}

\begin{claim}
    \label{c:complexity}
    The insert, delete and query operations run in time $\mathrm{O}(n)$, where $n$ is the number of BST nodes.
\end{claim}

The claim above follows from the fact that all rules match and apply in constant time, and that the BST is acyclic so that the loop \texttt{\{go\_right1, go\_left1\}!} always terminates after at most $n$ iterations, where $n$ is the number of BST nodes.

\input{rules_2}

\section{Runtime Measurements}

We provide runtime measurements of the program \texttt{bst}\footnote{Concrete syntax available at: \url{https://github.com/UoYCS-plasma/GP2/blob/master/programs/bst.gp2}.}
 in Figures \ref{fig:degen-plot} and \ref{fig:balanced-plot} (obtained using a quad-core Intel i5 clocked at 2.4 GHz,
with 16 GB RAM, running 64-bit Ubuntu 22.04). In Figure \ref{fig:degen-plot}, only degenerate trees (that is, linked lists) are considered, and the time required to perform a single operation is recorded. For each tree size, the operation was executed 300 times, and the results were averaged. Each operation takes longer on a linked list because performing any update or search requires traversal of all existing vertices. In contrast, on a balanced tree of the same size, such operations require visiting at most $\log n$ vertices, which is significantly fewer. This difference explains why operations are much faster in the second plot. In Figure \ref{fig:balanced-plot}, only fully balanced trees are considered. Each measurement records the time required to insert an element into a leaf of the tree, search for it, and subsequently delete it. As the runtime of each individual operation on balanced trees was too small to measure reliably, we consider the three operations together to obtain a more stable and representative measurement, and as with the degenerate trees, each sequence of operations was repeated 300 times, and the results were averaged.

\input{Plots/degen-plot}
\input{Plots/balanced-plot}

\section{Conclusion and Future Work}

We have shown how to implement binary search trees in the rule-based language GP\,2 such that searching, inserting, and deleting entries requires linear time in the worst-case. The next step will be to establish, both experimentally and analytically, that the operations run in time $\mathrm{O}(\log n)$ on average. 

Another topic for future work is to implement some form of \emph{balanced}\/ binary search trees, such as red–black trees, where the operations preserve balance. For such a structure, the query, insert, and delete operations are guaranteed to run in time $\mathrm{O}(\log n)$. An implementation of red–black trees in the graph transformation language PROGRES is discussed in \cite{A-E-H-H-K-K-P-S-T98a}, where the height of the left and right subtrees of a node are allowed to differ up to a factor of 2. We also mention \cite{Baldan-Corradini-Esparza-Heindel-Koenig-Kozioura05a} as a graph transformation approach to formally verify the insertion operation of red–black trees.

An alternative approach to data structures in GP\,2 could be to efficiently implement them in an imperative language such as C and provide access operations in GP\,2. However, we argue that for graph structures such as binary search trees, the code of operations is easier to read and understand in a rule-based graph transformation language. Moreover, writing both application programs and data structure implementations in GP \,2 allows formal program verification with, for example, the methods described in \cite{Poskitt-Plump23a,Wulandari-Plump21a}.

\bibliographystyle{eptcs}
\bibliography{generic}
\appendix

\end{document}

%% file: program.tex
\begin{figure}[!ht]

\begin{mdframed}[linewidth=1pt]
\begin{verbatim}
Main   = make_root; (try insert then Insert else (
                     try search then Search else (
                     try delete then Delete else fail)); 
                     try next_op else break)!
Insert = try root  then ({go_right1, go_left1}!; 
                   if match then break else add_leaf) else add_root
Search = try root  else break; {go_right1, go_left1}!; try match
Delete = try root  else break; {go_right1, go_left1}!;
         try Case1 else (try Case2 else Case3)
Case1  = try delete_leaf else fail
Case2  = try delete_midl else fail
Case3  = save_node; go_left2; if go_right2 
         then (go_right2!; {swap1, swap2, swap3, swap4}) 
         else {swap5, swap6}; try Case1 else Case2
         
\end{verbatim}
\begin{tikzpicture}
    \tikzstyle{every node}=[font=\ttfamily]
    \draw (0.65,0.75) node[align=left] {insert(n:int)};
    \node[rectangle, thick, rounded corners=7, draw=black, double, double distance=1pt,minimum size=5mm,
    label=below:\tiny\tiny1](a1) at (0,0){"i":n};
    \draw (1.5,0) node[] {$\Rightarrow$};
    \node[rectangle, thick, rounded corners=7, draw=black, double, double distance=1pt,minimum size=5mm,
    label=below:\tiny\tiny1](a2) at (3,0){"i":n};
    \draw (4,1) -- (4,-0.5);
\end{tikzpicture}
\hspace{0.5em}
\begin{tikzpicture}
    \tikzstyle{every node}=[font=\ttfamily]
    \draw (0.65,0.75) node[align=left] {search(n:int)};
    \node[rectangle, thick, rounded corners=7, draw=black, double, double distance=1pt,minimum size=5mm,
    label=below:\tiny\tiny1](a1) at (0,0){"s":n};
    \draw (1.5,0) node[] {$\Rightarrow$};
    \node[rectangle, thick, rounded corners=7, draw=black, double, double distance=1pt,minimum size=5mm,
    label=below:\tiny\tiny1](a2) at (3,0){"s":n};
    \draw (4,1) -- (4,-0.5);
\end{tikzpicture}
\hspace{0.5em}
\begin{tikzpicture}
    \tikzstyle{every node}=[font=\ttfamily]
    \draw (0.65,0.75) node[align=left] {delete(n:int)};
    \node[rectangle, thick, rounded corners=7, draw=black, double, double distance=1pt,minimum size=5mm,
    label=below:\tiny\tiny1](a1) at (0,0){"d":n};
    \draw (1.5,0) node[] {$\Rightarrow$};
    \node[rectangle, thick, rounded corners=7, draw=black, double, double distance=1pt,minimum size=5mm,
    label=below:\tiny\tiny1](a2) at (3,0){"d":n};
\end{tikzpicture}
\hspace{0.5em}

\begin{tikzpicture}
    \tikzstyle{every node}=[font=\ttfamily]
    \draw (1.3,0.75) node[align=left] {next_op(x,y:list)};
    \node[rectangle, thick, rounded corners=7, draw=black, double, double distance=1pt,minimum size=5mm,
    label=below:\tiny\tiny1](a1) at (0,0){x};
    \node[rectangle, thick, rounded corners=7, draw=black, minimum size=5mm,
    label=below:\tiny\tiny2](b1) at (1,0){y};
    \draw (2,0) node[] {$\Rightarrow$};
    \node[rectangle, thick, rounded corners=7, draw=black, minimum size=5mm,
    label=below:\tiny\tiny1](a2) at (3,0){x};
    \node[rectangle, thick, rounded corners=7, draw=black, double, double distance=1pt,minimum size=5mm,
    label=below:\tiny\tiny2](b2) at (4,0){y};
    \draw[very thick, ->] (a1) edge (b1) (a2) edge (b2);
    \draw (4.7,1) -- (4.7,-0.5);
\end{tikzpicture}
\hspace{0.5em}
\begin{tikzpicture}
    \tikzstyle{every node}=[font=\ttfamily]
    \draw (0.75,0.75) node[align=left] {make\_root()};
    \node[](a1) at (0,0){$\emptyset$};
    \draw (1,0) node[] {$\Rightarrow$};
    \node[rectangle, thick, rounded corners=7, draw=black, minimum size=5mm, fill=green, label=below:\tiny\tiny](a2) at (2,0){};
    \draw (0,-0.62) -- (0,-0.62);
    \draw (3,1) -- (3,-0.5);
\end{tikzpicture}
\hspace{0.5em}
\begin{tikzpicture}
    \tikzstyle{every node}=[font=\ttfamily]
    \draw (0.8,0.75) node[align=left] {root(a:int)};
    \node[rectangle, thick, rounded corners=7, draw=black, minimum size=5mm, fill=green, label=below:\tiny\tiny1](a1) at (0,0){};
    \node[rectangle, thick, rounded corners=7, draw=black, minimum size=5mm, fill=gray!60, label=below:\tiny\tiny2](a2) at (1,0){a};
    \draw (2,0) node[] {$\Rightarrow$};
    \node[rectangle, thick, rounded corners=7, draw=black, minimum size=5mm, fill=green, label=below:\tiny\tiny1](a3) at (3,0){};
    \node[rectangle, thick, rounded corners=7, draw=black, minimum size=5mm, double, double distance=1pt, fill=gray!60, label=below:\tiny\tiny2](a4) at (4,0){a};
    \draw[very thick, ->] (a1) edge (a2) (a3) edge (a4);
\end{tikzpicture}

\begin{tikzpicture}
    \tikzstyle{every node}=[font=\ttfamily]
    \draw (1.2,0.75) node[align=left] {add\_root(a:int)};
    \node[rectangle, thick, rounded corners=7, draw=black, minimum size=5mm, fill=green, label=below:\tiny\tiny1](a1) at (0,0){};

    \draw (2,-0.5) node[] {$\Rightarrow$};
    \node[rectangle, thick, rounded corners=7, draw=black, minimum size=5mm, fill=green, label=below:\tiny\tiny1](a3) at (3,0){};
    \node[rectangle, thick, rounded corners=7, draw=black, minimum size=5mm, double, double distance=1pt, label=below:\tiny\tiny2](b1) at (0.5,-1){"i":a};
    \node[rectangle, thick, rounded corners=7, draw=black, minimum size=5mm, fill=gray!60, label=below:\tiny\tiny](a4) at (4,0){a};
    \node[rectangle, thick, rounded corners=7, draw=black, minimum size=5mm, double, double distance=1pt, label=below:\tiny\tiny2](b2) at (3.5,-1){"i":a};
    \draw[very thick, ->] (a3) edge (a4);
    \draw (4.8,1) -- (4.8,-1.5);
\end{tikzpicture}
\hspace{0.5em}
\begin{tikzpicture}
    \tikzstyle{every node}=[font=\ttfamily]
    \draw (1.4,0.75) node[align=left] {add\_leaf(a,x:int)};
    \node[rectangle, thick, rounded corners=7, draw=black, minimum size=5mm, double, double distance=1pt,fill=gray!60, label=below:\tiny\tiny1](a1) at (0,0){x};

    \draw (2,-0.5) node[] {$\Rightarrow$};
    \node[rectangle, thick, rounded corners=7, draw=black, minimum size=5mm,fill=gray!60, label=below:\tiny\tiny1](a3) at (3,0){x};
    \node[rectangle, thick, rounded corners=7, draw=black, minimum size=5mm, double, double distance=1pt, label=below:\tiny\tiny2](b1) at (0.5,-1){"i":a};
    \node[rectangle, thick, rounded corners=7, draw=black, minimum size=5mm, fill=gray!60, label=below:\tiny\tiny](a4) at (4,0){a};
    \node[rectangle, thick, rounded corners=7, draw=black, minimum size=5mm, double, double distance=1pt, label=below:\tiny\tiny2](b2) at (3.5,-1){"i":a};
    \draw[very thick, ->] (a3) edge (a4);
    \draw (4.8,1) -- (4.8,-1.5);
\end{tikzpicture}
\hspace{0.5em}
\begin{tikzpicture}
    \tikzstyle{every node}=[font=\ttfamily]
    \draw (1.4,0.75) node[align=left] {match(o:char;a:int)};
    \node[rectangle, thick, rounded corners=7, draw=black, minimum size=5mm, double, double distance=1pt,fill=gray!60, label=below:\tiny\tiny1](a1) at (0,0){a};

    \draw (1,-0.5) node[] {$\Rightarrow$};
    \node[rectangle, thick, rounded corners=7, draw=black, minimum size=5mm, double, double distance=1pt, label=below:\tiny\tiny2](b1) at (0,-1){o:a};
    \node[rectangle, thick, rounded corners=7, draw=black, minimum size=5mm, double, double distance=1pt,fill=gray!60, label=below:\tiny\tiny1](a4) at (2,0){a};
    \node[rectangle, thick, rounded corners=7, draw=black, minimum size=5mm, double, double distance=1pt, label=below:\tiny\tiny2](b2) at (2,-1){o:a};
    \draw[very thick, ->] (b2) edge[dashed, bend left] (a4);
\end{tikzpicture}
\\
\begin{tikzpicture}
    \tikzstyle{every node}=[font=\ttfamily]
    \draw (2.4,0.75) node[align=left] {go\_right1(o:char;x,n,m:int)};
    \node[rectangle, thick, rounded corners=7, draw=black,double, double distance=1pt, minimum size=5mm, fill=gray!60, label=below:\tiny\tiny1](a2) at (0,0){n};
    \node[rectangle, thick, rounded corners=7, draw=black, minimum size=5mm, fill=gray!60, label=below:\tiny\tiny2](a1) at (1,0){m};

    \draw (2,-0.5) node[] {$\Rightarrow$};
    \node[rectangle, thick, rounded corners=7, draw=black, minimum size=5mm, fill=gray!60, label=below:\tiny\tiny1](a3) at (3,0){n};
    \node[rectangle, thick, rounded corners=7, draw=black, minimum size=5mm, double, double distance=1pt, label=below:\tiny\tiny3](b1) at (0.5,-1){o:x};
    \node[rectangle, thick, rounded corners=7, draw=black, double, double distance=1pt, minimum size=5mm, fill=gray!60, label=below:\tiny\tiny2](a4) at (4,0){m};
    \node[rectangle, thick, rounded corners=7, draw=black, minimum size=5mm, double, double distance=1pt, label=below:\tiny\tiny3](b2) at (3.5,-1){o:x};
    \draw[very thick, ->] (a3) edge (a4);
    \draw[very thick, ->] (a2) edge (a1);
    \draw (1.95,-1.9) node[align=left] {where(m > n and x > n)};
    \draw (5.4,1) -- (5.4,-2);
\end{tikzpicture}
\hspace{0.5em}
\begin{tikzpicture}
    \tikzstyle{every node}=[font=\ttfamily]
    \draw (2.35,0.75) node[align=left] {go\_left1(o:char;x,n,m:int)};
    \node[rectangle, thick, rounded corners=7, draw=black,double, double distance=1pt, minimum size=5mm, fill=gray!60, label=below:\tiny\tiny1](a2) at (0,0){n};
    \node[rectangle, thick, rounded corners=7, draw=black, minimum size=5mm, fill=gray!60, label=below:\tiny\tiny2](a1) at (1,0){m};

    \draw (2,-0.5) node[] {$\Rightarrow$};
    \node[rectangle, thick, rounded corners=7, draw=black, minimum size=5mm, fill=gray!60, label=below:\tiny\tiny1](a3) at (3,0){n};
    \node[rectangle, thick, rounded corners=7, draw=black, minimum size=5mm, double, double distance=1pt, label=below:\tiny\tiny3](b1) at (0.5,-1){o:x};
    \node[rectangle, thick, rounded corners=7, draw=black, double, double distance=1pt, minimum size=5mm, fill=gray!60, label=below:\tiny\tiny2](a4) at (4,0){m};
    \node[rectangle, thick, rounded corners=7, draw=black, minimum size=5mm, double, double distance=1pt, label=below:\tiny\tiny3](b2) at (3.5,-1){o:x};
    \draw[very thick, ->] (a3) edge (a4);
    \draw[very thick, ->] (a2) edge (a1);
    \draw (1.95,-1.9) node[align=left] {where(m < n and x < n)};
\end{tikzpicture}
\\
\begin{tikzpicture}
    \tikzstyle{every node}=[font=\ttfamily]
    \draw (1.45,0.75) node[align=left] {go\_left2(n,m:int)};
    \node[rectangle, thick, rounded corners=7, draw=black,double, double distance=1pt, minimum size=5mm, fill=gray!60, label=below:\tiny\tiny1](a2) at (0,0){n};
    \node[rectangle, thick, rounded corners=7, draw=black, minimum size=5mm, fill=gray!60, label=below:\tiny\tiny2](a1) at (1,0){m};

    \draw (2,0) node[] {$\Rightarrow$};
    \node[rectangle, thick, rounded corners=7, draw=black, minimum size=5mm, fill=gray!60, label=below:\tiny\tiny1](a3) at (3,0){n};
    \node[rectangle, thick, rounded corners=7, draw=black, double, double distance=1pt, minimum size=5mm, fill=gray!60, label=below:\tiny\tiny2](a4) at (4,0){m};
    \draw[very thick, ->] (a3) edge (a4);
    \draw[very thick, ->] (a2) edge (a1);
    \draw (0.9,-1.0) node[align=left] {where(m < n)};
    \draw (4.8,1) -- (4.8,-1.2);
\end{tikzpicture}
\hspace{0.5em}
\begin{tikzpicture}
    \tikzstyle{every node}=[font=\ttfamily]
    \draw (1.5,0.75) node[align=left] {go\_right2(n,m:int)};
    \node[rectangle, thick, rounded corners=7, draw=black,double, double distance=1pt, minimum size=5mm, fill=gray!60, label=below:\tiny\tiny1](a2) at (0,0){n};
    \node[rectangle, thick, rounded corners=7, draw=black, minimum size=5mm, fill=gray!60, label=below:\tiny\tiny2](a1) at (1,0){m};

    \draw (2,0) node[] {$\Rightarrow$};
    \node[rectangle, thick, rounded corners=7, draw=black, minimum size=5mm, fill=gray!60, label=below:\tiny\tiny1](a3) at (3,0){n};
    \node[rectangle, thick, rounded corners=7, draw=black, double, double distance=1pt, minimum size=5mm, fill=gray!60, label=below:\tiny\tiny2](a4) at (4,0){m};
    \draw[very thick, ->] (a3) edge (a4);
    \draw[very thick, ->] (a2) edge (a1);
    \draw (0.9,-1.0) node[align=left] {where(m > n)};
\end{tikzpicture}

\end{mdframed}
\caption{The program \texttt{bst}; the rest of the rules are graphically depicted in Figure~\ref{fig:bst-prog2}.}
\label{fig:bst-prog}
\end{figure}
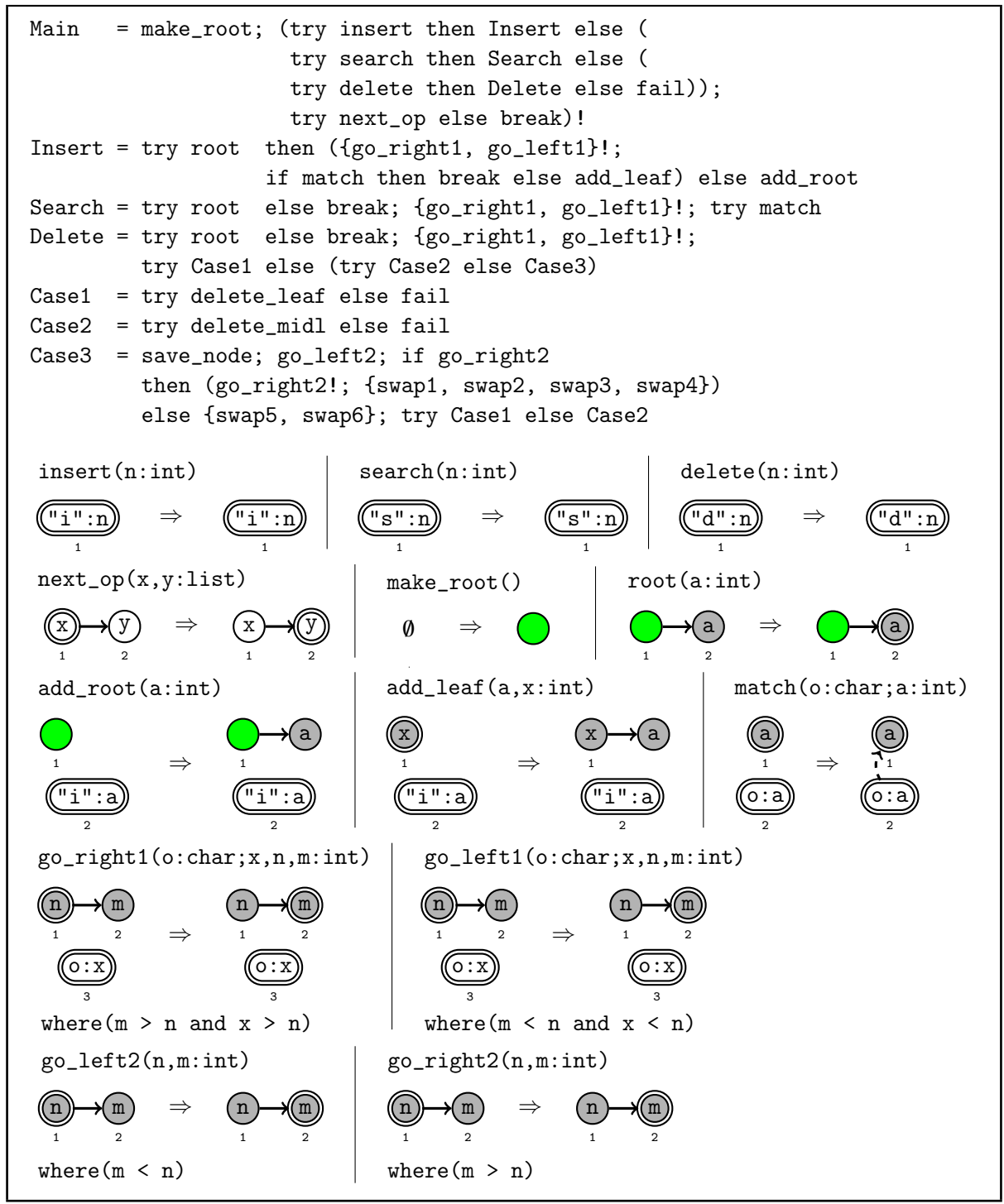

%% file: rules_2.tex
\begin{figure}[!ht]

\begin{mdframed}[linewidth=1pt]
\begin{tikzpicture}
    \tikzstyle{every node}=[font=\ttfamily]
    \draw (1.8,0.75) node[align=left] {delete\_leaf(x,y:list)};
    \node[rectangle, thick, rounded corners=7, draw=black, minimum size=5mm, fill=magenta!60, label=below:\tiny\tiny1](a2) at (0,0){x};
    \node[rectangle, thick, rounded corners=7, draw=black,double, double distance=1pt, minimum size=5mm, fill=gray!60, label=below:\tiny\tiny2](a1) at (1,0){y};

    \draw (2,0) node[] {$\Rightarrow$};
    \node[rectangle, thick, rounded corners=7, draw=black, minimum size=5mm, fill=magenta!60, label=below:\tiny\tiny1](a3) at (3,0){x};
    \node[rectangle, thick, rounded corners=7, draw=black, double, double distance=1pt, minimum size=5mm, fill=gray!60, label=below:\tiny\tiny2](a4) at (4,0){y};
    \draw[very thick, ->] (a2) edge (a1);
    \draw (1.7,-1.0) node[align=left] {where(outdeg(2) = 0)};
    \draw (4.8,1) -- (4.8,-1.2);
\end{tikzpicture}
\hspace{0.5em}
\begin{tikzpicture}
    \tikzstyle{every node}=[font=\ttfamily]
    \draw (2,0.75) node[align=left] {delete\_midl(x,y,z:list)};
    \node[rectangle, thick, rounded corners=7, draw=black, minimum size=5mm, fill=magenta!60, label=below:\tiny\tiny1](a2) at (0,0){x};
    \node[rectangle, thick, rounded corners=7, draw=black,double, double distance=1pt, minimum size=5mm, fill=gray!60, label=below:\tiny\tiny2](a1) at (1,0){y};
    \node[rectangle, thick, rounded corners=7, draw=black, minimum size=5mm, fill=gray!60, label=below:\tiny\tiny3](a3) at (2,0){z};

    \draw (3,0) node[] {$\Rightarrow$};
    \node[rectangle, thick, rounded corners=7, draw=black, minimum size=5mm, fill=magenta!60, label=below:\tiny\tiny1](a4) at (4,0){x};
    \node[rectangle, thick, rounded corners=7, draw=black, double, double distance=1pt, minimum size=5mm, fill=gray!60, label=below:\tiny\tiny2](a5) at (5,0){y};
    \node[rectangle, thick, rounded corners=7, draw=black, minimum size=5mm, fill=gray!60, label=below:\tiny\tiny3](a6) at (6,0){y};
    \draw[very thick, ->] (a2) edge (a1);
    \draw[very thick, ->] (a1) edge (a3);
    \draw[very thick, ->] (a4) edge[bend left = 40] (a6);
    \draw (1.8,-1.0) node[align=left] {where(outdeg(2) = 1)};
\end{tikzpicture}
\\
\begin{tikzpicture}
    \tikzstyle{every node}=[font=\ttfamily]
    \draw (1.4,0.75) node[align=left] {save\_node(x:list)};
    \node[rectangle, thick, rounded corners=7, draw=black, double, double distance=1pt, minimum size=5mm, fill=gray!60, label=below:\tiny\tiny1](a1) at (0,0){x};

    \draw (1,0) node[] {$\Rightarrow$};
    \node[rectangle, thick, rounded corners=7, draw=black, double, double distance=1pt, minimum size=5mm, fill=gray!60, label=below:\tiny\tiny1](a3) at (2,0){x};
    \node[rectangle, thick, rounded corners=7, draw=black, double, double distance=1pt, minimum size=5mm, fill=red!60, label=below:\tiny\tiny](a4) at (3,0){};
    \draw[very thick, ->] (a4) edge[red] (a3);
\end{tikzpicture}
\\
\begin{tikzpicture}
    \tikzstyle{every node}=[font=\ttfamily]
    \draw (2.25,0.75) node[align=left] {swap1(a,b,c,d,e,f,g:list)};
    \node[rectangle, thick, rounded corners=7, draw=black, minimum size=5mm, fill=gray!60, label=below:\tiny\tiny1](a) at (0,0){a};
    \node[rectangle, thick, rounded corners=7, draw=black, minimum size=5mm, double, double distance=1pt,fill=gray!60, label=below:\tiny\tiny2](b) at (1.6,-0.5){b};
    \node[rectangle, thick, rounded corners=7, draw=black, minimum size=5mm, fill=gray!60, label=below:\tiny\tiny3](c) at (0,-1){c};

    \node[rectangle, thick, rounded corners=7, draw=black, minimum size=5mm, fill=magenta!60, label=below:\tiny\tiny4](d) at (2.8,0){d};
    \node[rectangle, thick, rounded corners=7, draw=black, minimum size=5mm, fill=gray!60, label=below:\tiny\tiny5](e) at (3.5,-0.5){e};
    \node[rectangle, thick, rounded corners=7, draw=black, minimum size=5mm, fill=gray!60, label=below:\tiny\tiny6](f) at (2.8,-1){f};
    \node[rectangle, thick, rounded corners=7, draw=black, minimum size=5mm, fill=gray!60, label=below:\tiny\tiny7](g) at (4.2,-1){g};
    \node[rectangle, thick, rounded corners=7, draw=black, double, double distance = 1pt, minimum size=5mm, fill=red!60, label=below:\tiny\tiny](r) at (4.2,0){};

    \draw[very thick, ->] (a) edge (b) (b) edge (c);
    \draw[very thick, ->] (d) edge[bend right] (e) (r) edge[red, bend left] (e);
    \draw[very thick, ->] (d) edge[bend right] (e) (e) edge[bend left] (f) (e) edge[bend right] (g);
    
    \draw (5.6,-0.5) node[] {$\Rightarrow$};

    \node[rectangle, thick, rounded corners=7, draw=black, minimum size=5mm, fill=gray!60, label=below:\tiny\tiny1](a1) at (7,0){a};
    \node[rectangle, thick, rounded corners=7, draw=black, minimum size=5mm, double, double distance=1pt,fill=gray!60, label=below:\tiny\tiny5](b1) at (8.6,-0.5){e};
    \node[rectangle, thick, rounded corners=7, draw=black, minimum size=5mm, fill=gray!60, label=below:\tiny\tiny3](c1) at (7,-1){c};

    \node[rectangle, thick, rounded corners=7, draw=black, minimum size=5mm, fill=magenta!60, label=below:\tiny\tiny4](d1) at (9.8,0){d};
    \node[rectangle, thick, rounded corners=7, draw=black, minimum size=5mm, fill=gray!60, label=below:\tiny\tiny2](e1) at (10.5,-0.5){b};
    \node[rectangle, thick, rounded corners=7, draw=black, minimum size=5mm, fill=gray!60, label=below:\tiny\tiny6](f1) at (9.8,-1){f};
    \node[rectangle, thick, rounded corners=7, draw=black, minimum size=5mm, fill=gray!60, label=below:\tiny\tiny7](g1) at (11.2,-1){g};

    \draw[very thick, ->] (a1) edge (b1) (b1) edge (c1);
    \draw[very thick, ->] (d1) edge[bend right] (e1);
    \draw[very thick, ->] (d1) edge[bend right] (e1) (e1) edge[bend left] (f1) (e1) edge[bend right] (g1);
    
    \draw (0,-1.2) -- (0,-1.2);
\end{tikzpicture}

\begin{tikzpicture}
    \tikzstyle{every node}=[font=\ttfamily]
    \draw (2,0.75) node[align=left] {swap2(a,b,d,e,f,g:list)};
    \draw (1.7,-1.9) node[align=left] {where(outdeg(2) = 0)};
    \node[rectangle, thick, rounded corners=7, draw=black, minimum size=5mm, fill=gray!60, label=below:\tiny\tiny1](a) at (0,0){a};
    \node[rectangle, thick, rounded corners=7, draw=black, minimum size=5mm, double, double distance=1pt,fill=gray!60, label=below:\tiny\tiny2](b) at (1.6,-0.25){b};

    \node[rectangle, thick, rounded corners=7, draw=black, minimum size=5mm, fill=magenta!60, label=below:\tiny\tiny4](d) at (2.8,0){d};
    \node[rectangle, thick, rounded corners=7, draw=black, minimum size=5mm, fill=gray!60, label=below:\tiny\tiny5](e) at (3.5,-0.5){e};
    \node[rectangle, thick, rounded corners=7, draw=black, minimum size=5mm, fill=gray!60, label=below:\tiny\tiny6](f) at (2.8,-1){f};
    \node[rectangle, thick, rounded corners=7, draw=black, minimum size=5mm, fill=gray!60, label=below:\tiny\tiny7](g) at (4.2,-1){g};
    \node[rectangle, thick, rounded corners=7, draw=black, minimum size=5mm, double, double distance = 1pt, fill=red!60, label=below:\tiny\tiny](r) at (4.2,0){};

    \draw[very thick, ->] (a) edge (b);
    \draw[very thick, ->] (d) edge[bend right] (e) (r) edge[red, bend left] (e);
    \draw[very thick, ->] (d) edge[bend right] (e) (e) edge[bend left] (f) (e) edge[bend right] (g);
    
    \draw (5.6,-0.5) node[] {$\Rightarrow$};

    \node[rectangle, thick, rounded corners=7, draw=black, minimum size=5mm, fill=gray!60, label=below:\tiny\tiny1](a1) at (7,0){a};
    \node[rectangle, thick, rounded corners=7, draw=black, minimum size=5mm, double, double distance=1pt,fill=gray!60, label=below:\tiny\tiny5](b1) at (8.6,-0.25){e};

    \node[rectangle, thick, rounded corners=7, draw=black, minimum size=5mm, fill=magenta!60, label=below:\tiny\tiny4](d1) at (9.8,0){d};
    \node[rectangle, thick, rounded corners=7, draw=black, minimum size=5mm, fill=gray!60, label=below:\tiny\tiny2](e1) at (10.5,-0.5){b};
    \node[rectangle, thick, rounded corners=7, draw=black, minimum size=5mm, fill=gray!60, label=below:\tiny\tiny6](f1) at (9.8,-1){f};
    \node[rectangle, thick, rounded corners=7, draw=black, minimum size=5mm, fill=gray!60, label=below:\tiny\tiny7](g1) at (11.2,-1){g};

    \draw[very thick, ->] (a1) edge (b1);
    \draw[very thick, ->] (d1) edge[bend right] (e1);
    \draw[very thick, ->] (d1) edge[bend right] (e1) (e1) edge[bend left] (f1) (e1) edge[bend right] (g1);
    
    \draw (0,-1.2) -- (0,-1.2);
\end{tikzpicture}

\begin{tikzpicture}
    \tikzstyle{every node}=[font=\ttfamily]
    \draw (2,0.75) node[align=left] {swap3(a,b,c,d,e,f:list)};

    \node[rectangle, thick, rounded corners=7, draw=black, minimum size=5mm, fill=magenta!60, label=below:\tiny\tiny1](a) at (0,0){a};
    \node[rectangle, thick, rounded corners=7, draw=black, minimum size=5mm, fill=gray!60, label=below:\tiny\tiny2](b) at (0.7,-0.5){b};
    \node[rectangle, thick, rounded corners=7, draw=black, minimum size=5mm, fill=gray!60, label=below:\tiny\tiny3](c) at (0,-1){c};
    \node[rectangle, thick, rounded corners=7, draw=black, minimum size=5mm, fill=gray!60, label=below:\tiny\tiny4](d) at (1.4,-1){d};
    \node[rectangle, thick, rounded corners=7, draw=black, minimum size=5mm, double, double distance = 1pt, fill=gray!60, label=below:\tiny\tiny5](e) at (0.7,-1.5){e};
    \node[rectangle, thick, rounded corners=7, draw=black, minimum size=5mm, fill=gray!60, label=below:\tiny\tiny6](f) at (0,-2.2){f};
    \node[rectangle, thick, rounded corners=7, draw=black, minimum size=5mm, double, double distance = 1pt, fill=red!60, label=below:\tiny\tiny](r) at (1.4,0){};

    \draw[very thick, ->] (a) edge[bend right] (b) (r) edge[red, bend left] (b) (b) edge[bend left] (c) (b) edge[bend right] (d) (c) edge[bend right] (e) (e) edge[bend left] (f);
    
    \draw (2.8,-1) node[] {$\Rightarrow$};

    \node[rectangle, thick, rounded corners=7, draw=black, minimum size=5mm, fill=magenta!60, label=below:\tiny\tiny1](a1) at (4.2,0){a};
    \node[rectangle, thick, rounded corners=7, draw=black, minimum size=5mm, fill=gray!60, label=below:\tiny\tiny5](b1) at (4.9,-0.5){e};
    \node[rectangle, thick, rounded corners=7, draw=black, minimum size=5mm, fill=gray!60, label=below:\tiny\tiny3](c1) at (4.2,-1){c};
    \node[rectangle, thick, rounded corners=7, draw=black, minimum size=5mm, fill=gray!60, label=below:\tiny\tiny4](d1) at (5.6,-1){d};
    \node[rectangle, thick, rounded corners=7, draw=black, minimum size=5mm, double, double distance = 1pt, fill=gray!60, label=below:\tiny\tiny2](e1) at (4.9,-1.5){b};
    \node[rectangle, thick, rounded corners=7, draw=black, minimum size=5mm, fill=gray!60, label=below:\tiny\tiny6](f1) at (4.2,-2.2){f};

    \draw[very thick, ->] (a1) edge[bend right] (b1)  (b1) edge[bend left] (c1) (b1) edge[bend right] (d1) (c1) edge[bend right] (e1) (e1) edge[bend left] (f1);
    
    \draw (7.1,1) -- (7.1,-2.8);
    \draw (0,-1.2) -- (0,-1.2);
\end{tikzpicture}
\hspace{0.5em}
\begin{tikzpicture}
    \tikzstyle{every node}=[font=\ttfamily]
    \draw (1.8,0.75) node[align=left] {swap4(a,b,c,d,e:list)};

    \node[rectangle, thick, rounded corners=7, draw=black, minimum size=5mm, fill=magenta!60, label=below:\tiny\tiny1](a) at (0,0){a};
    \node[rectangle, thick, rounded corners=7, draw=black, minimum size=5mm, fill=gray!60, label=below:\tiny\tiny2](b) at (0.7,-0.5){b};
    \node[rectangle, thick, rounded corners=7, draw=black, minimum size=5mm, fill=gray!60, label=below:\tiny\tiny3](c) at (0,-1){c};
    \node[rectangle, thick, rounded corners=7, draw=black, minimum size=5mm, fill=gray!60, label=below:\tiny\tiny4](d) at (1.4,-1){d};
    \node[rectangle, thick, rounded corners=7, draw=black, minimum size=5mm, double, double distance = 1pt, fill=gray!60, label=below:\tiny\tiny5](e) at (0.7,-1.5){e};
    \node[rectangle, thick, rounded corners=7, draw=black, minimum size=5mm, double, double distance = 1pt, fill=red!60, label=below:\tiny\tiny](r) at (1.4,0){};

    \draw[very thick, ->] (a) edge[bend right] (b) (r) edge[red, bend left] (b) (b) edge[bend left] (c) (b) edge[bend right] (d) (c) edge[bend right] (e);
    
    \draw (2.8,-1) node[] {$\Rightarrow$};

    \node[rectangle, thick, rounded corners=7, draw=black, minimum size=5mm, fill=magenta!60, label=below:\tiny\tiny1](a1) at (4.2,0){a};
    \node[rectangle, thick, rounded corners=7, draw=black, minimum size=5mm, fill=gray!60, label=below:\tiny\tiny5](b1) at (4.9,-0.5){e};
    \node[rectangle, thick, rounded corners=7, draw=black, minimum size=5mm, fill=gray!60, label=below:\tiny\tiny3](c1) at (4.2,-1){c};
    \node[rectangle, thick, rounded corners=7, draw=black, minimum size=5mm, fill=gray!60, label=below:\tiny\tiny4](d1) at (5.6,-1){d};
    \node[rectangle, thick, rounded corners=7, draw=black, minimum size=5mm, double, double distance = 1pt, fill=gray!60, label=below:\tiny\tiny2](e1) at (4.9,-1.5){b};

    \draw[very thick, ->] (a1) edge[bend right] (b1)  (b1) edge[bend left] (c1) (b1) edge[bend right] (d1) (c1) edge[bend right] (e1);

    \draw (1.75,-2.56) node[align=left] {where(outdeg(5) = 0)};
    
    \draw (0,-1.2) -- (0,-1.2);
\end{tikzpicture}

\begin{tikzpicture}
    \tikzstyle{every node}=[font=\ttfamily]
    \draw (1.2,0.75) node[align=left] {swap5(a,b,c,d,e:list)};

    \node[rectangle, thick, rounded corners=7, draw=black, minimum size=5mm, fill=magenta!60, label=below:\tiny\tiny1](a) at (0,0){a};
    \node[rectangle, thick, rounded corners=7, draw=black, minimum size=5mm, fill=gray!60, label=below:\tiny\tiny2](b) at (0.7,-0.5){b};
    \node[rectangle, thick, rounded corners=7, draw=black, minimum size=5mm, fill=gray!60, double, double distance = 1pt, label=below:\tiny\tiny3](c) at (0,-1){c};
    \node[rectangle, thick, rounded corners=7, draw=black, minimum size=5mm, fill=gray!60, label=below:\tiny\tiny4](d) at (1.4,-1){d};
    \node[rectangle, thick, rounded corners=7, draw=black, minimum size=5mm, fill=gray!60, label=below:\tiny\tiny5](e) at (-0.7,-1.5){e};
    \node[rectangle, thick, rounded corners=7, draw=black, minimum size=5mm, double, double distance = 1pt, fill=red!60, label=below:\tiny\tiny](r) at (1.4,0){};

    \draw[very thick, ->] (a) edge[bend right] (b) (r) edge[red, bend left] (b) (b) edge[bend left] (c) (b) edge[bend right] (d) (c) edge[bend left] (e);
    
    \draw (2.5,-1) node[] {$\Rightarrow$};

    \node[rectangle, thick, rounded corners=7, draw=black, minimum size=5mm, fill=magenta!60, label=below:\tiny\tiny1](a1) at (4.2,0){a};
    \node[rectangle, thick, rounded corners=7, draw=black, minimum size=5mm, fill=gray!60, label=below:\tiny\tiny3](b1) at (4.9,-0.5){c};
    \node[rectangle, thick, rounded corners=7, draw=black, minimum size=5mm, fill=gray!60, double, double distance = 1pt, label=below:\tiny\tiny2](c1) at (4.2,-1){b};
    \node[rectangle, thick, rounded corners=7, draw=black, minimum size=5mm, fill=gray!60, label=below:\tiny\tiny4](d1) at (5.6,-1){d};
    \node[rectangle, thick, rounded corners=7, draw=black, minimum size=5mm, fill=gray!60, label=below:\tiny\tiny5](e1) at (3.5,-1.5){e};

    \draw[very thick, ->] (a1) edge[bend right] (b1) (b1) edge[bend left] (c1) (b1) edge[bend right] (d1) (c1) edge[bend left] (e1);
    
    \draw (6.5,1) -- (6.5,-2.5);
    \draw (0,-1.2) -- (0,-1.2);
\end{tikzpicture}
\hspace{0.5em}
\begin{tikzpicture}
    \tikzstyle{every node}=[font=\ttfamily]
    \draw (1.62,0.75) node[align=left] {swap6(a,b,c,d:list)};

    \node[rectangle, thick, rounded corners=7, draw=black, minimum size=5mm, fill=magenta!60, label=below:\tiny\tiny1](a) at (0,0){a};
    \node[rectangle, thick, rounded corners=7, draw=black, minimum size=5mm, fill=gray!60, label=below:\tiny\tiny2](b) at (0.7,-0.5){b};
    \node[rectangle, thick, rounded corners=7, draw=black, minimum size=5mm, fill=gray!60, double, double distance = 1pt, label=below:\tiny\tiny3](c) at (0,-1){c};
    \node[rectangle, thick, rounded corners=7, draw=black, minimum size=5mm, fill=gray!60, label=below:\tiny\tiny4](d) at (1.4,-1){d};
    \node[rectangle, thick, rounded corners=7, draw=black, minimum size=5mm, double, double distance = 1pt, fill=red!60, label=below:\tiny\tiny](r) at (1.4,0){};

    \draw[very thick, ->] (a) edge[bend right] (b) (r) edge[red, bend left] (b) (b) edge[bend left] (c) (b) edge[bend right] (d);
    
    \draw (2.75,-0.5) node[] {$\Rightarrow$};

    \node[rectangle, thick, rounded corners=7, draw=black, minimum size=5mm, fill=magenta!60, label=below:\tiny\tiny1](a1) at (4.2,0){a};
    \node[rectangle, thick, rounded corners=7, draw=black, minimum size=5mm, fill=gray!60, label=below:\tiny\tiny3](b1) at (4.9,-0.5){c};
    \node[rectangle, thick, rounded corners=7, draw=black, minimum size=5mm, fill=gray!60, double, double distance = 1pt, label=below:\tiny\tiny2](c1) at (4.2,-1){b};
    \node[rectangle, thick, rounded corners=7, draw=black, minimum size=5mm, fill=gray!60, label=below:\tiny\tiny4](d1) at (5.6,-1){d};

    \draw[very thick, ->] (a1) edge[bend right] (b1) (b1) edge[bend left] (c1) (b1) edge[bend right] (d1);
    
    \draw (1.75,-2.2) node[align=left] {where(outdeg(3) = 0)};
    \draw (0,-1.2) -- (0,-1.2);
\end{tikzpicture}

\end{mdframed}
\caption{Remaining rules of program \texttt{bst}.}
\label{fig:bst-prog2}
\end{figure}
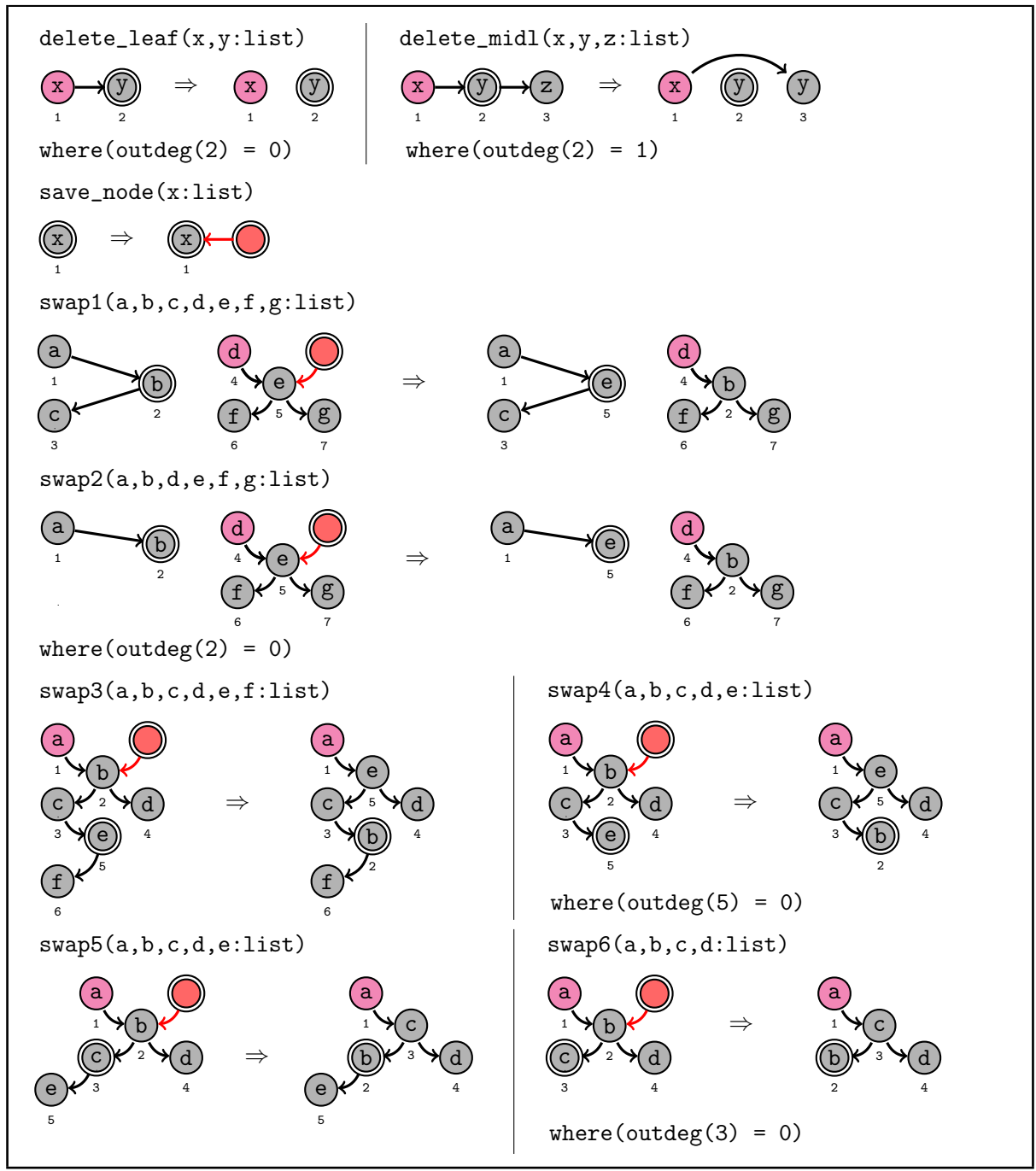

%% file: Plots/degen-plot.tex
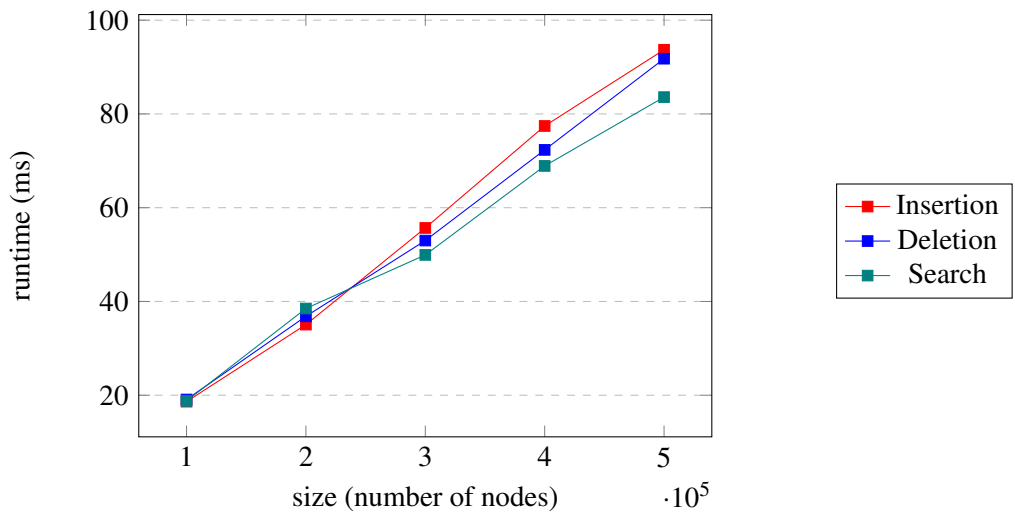
\begin{figure}[!ht]
    \centering
    \begin{tikzpicture}
    \begin{axis}[
      xlabel=size (number of nodes),
      ylabel=runtime (ms), ylabel style={above=0.2mm},
      width=9.2cm,height=7.2cm,
      legend style={at={(1.525,0.6)}},
      ymajorgrids=true,
      grid style=dashed]
      \addplot[color=red, mark=square*] table [y=time, x=n]{Data/degen-insert.dat};
      \addlegendentry{Insertion}
      \addplot[color=blue, mark=square*] table [y=time, x=n]{Data/degen-delete.dat};
      \addlegendentry{Deletion}
      \addplot[color=teal, mark=square*] table [y=time, x=n]{Data/degen-search.dat};
      \addlegendentry{Search}
    \end{axis}  
    \end{tikzpicture}
    \caption{Measured runtime per operation on degenerated trees.}
    \label{fig:degen-plot}
\end{figure}

%% file: Plots/balanced-plot.tex
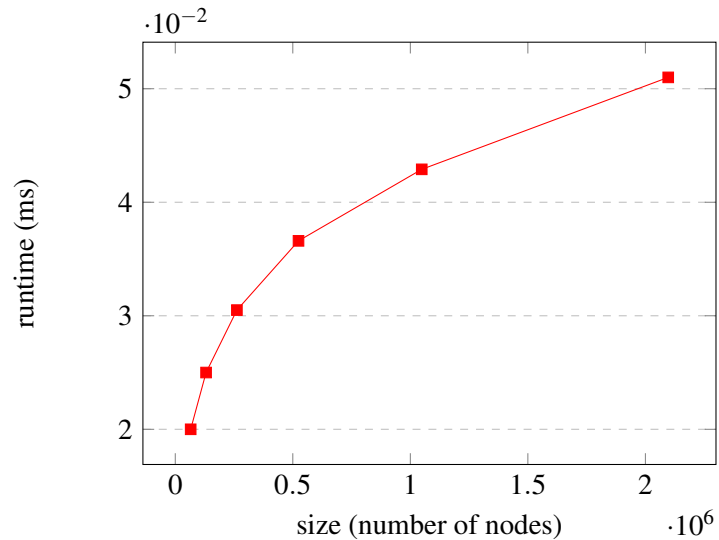
\begin{figure}[!ht]
    \centering
    \begin{tikzpicture}
    \begin{axis}[
      xlabel=size (number of nodes),
      ylabel=runtime (ms), ylabel style={above=0.2mm},
      width=9.2cm,height=7.2cm,
      legend style={at={(1.525,0.6)}},
      ymajorgrids=true,
      grid style=dashed]
      \addplot[color=red, mark=square*] table [y=time, x=n]{Data/balanced-insert.dat};
    \end{axis}  
    \end{tikzpicture}
    \caption{Measured averaged runtime for an insertion, followed by a search and deletion, on balanced trees.}
    \label{fig:balanced-plot}
\end{figure}